\documentclass[showpacs,amssymb,aps,nofootinbib,floatfix]{revtex4-1}
\usepackage{epsfig}
\newcommand{\be}{\begin{eqnarray}}
\newcommand{\ee}{\end{eqnarray}}
\usepackage{graphicx}
\newcommand{\ave}[1]{\left\langle #1 \right\rangle}
\newcommand{\order}[1]{ \mathcal{O} \left( #1 \right) }

  \newcommand{\lqcd}{\Lambda_{QCD}}

\begin{document} \hbadness=10000
\topmargin -0.8cm\oddsidemargin = -0.7cm\evensidemargin = -0.7cm
\title{The nuclear liquid-gas phase transition at large $N_c$ in the Van Der Waals approximation}
\author{Giorgio Torrieri, Igor Mishustin}
\affiliation{FIAS,
  J.W. Goethe Universit\"at, Max von Laue-Stra\ss{}e 1, 60438 Frankfurt am Main, Germany
torrieri@fias.uni-frankfurt.de}
\date{May 24, 2010}

\begin{abstract}
We examine the nuclear liquid-gas phase transition at large number of colors ($N_c$) within the framework of the Van Der Waals (VdW) model.   We argue that the VdW equation is appropriate at describing inter-nucleon forces , and discuss how each parameter scales with $N_c$.
We demonstrate that $N_c=3$ ( our world ) is not large with respect to the other dimensionless scale relevant to baryonic matter, the number of neighbors in a dense system $N_N$.
Consequently, we show that the liquid-gas phase transition looks dramatically different at $N_c \rightarrow \infty$ with respect of our world:  The critical point temperature becomes of the order of $\lqcd$ rather than below it.  The critical point density becomes of the order of the baryonic density, rather than an order of magnitude below it.   These are precisely the characteristics usually associated with the ``Quarkyonic phase''.   We therefore conjecture that quarkyonic matter is simply the large $N_c$ limit of the nuclear liquid, and the interplay between $N_c$ and $N_N$ is the reason why the nuclear liquid in our world is so different from quarkyonic matter.  We conclude by suggesting ways our conjecture can be tested in future lattice measurements.
\end{abstract}
\maketitle
\section{Introduction}
Strongly interacting matter at moderate ($\sim$ the confinement scale) quark chemical potential $\mu_q$ and moderate temperature $T$ has recently received a considerable amount of both theoretical and experimental interest.   Such matter can hopefully be produced in heavy ion collisions \cite{low1,low2,low3,low4}, and is thought to exhibit a rich phenomenology.  The latter includes one \cite{critical} or more \cite{ferroni} critical points, spinodal instabilities \cite{spino}, precursors to color superconductivity \cite{blas}, separation between chiral symmetry and confinement \cite{hansen,ratti,dirkq,sasaki1,sasaki2,sasaki3} chirally inhomogeneus phases \cite{chiralspiral1,chiralspiral2,buballa1,buballa2}, new phases \cite{quarkyonic} etc.

These conjectures are, however, extraordinarily difficult to quantitatively explore in a rigorous manner.
The quark chemical potential $\mu_q$ is nowhere near the asymptotic freedom limit where perturbative QCD can be used \cite{jansbook}.   It is, however, way too high for existing lattice-based approaches, dependent on $\mu_q/T \ll 1$, to work \cite{latticebook}.

\begin{figure*}
\includegraphics[scale=0.5]{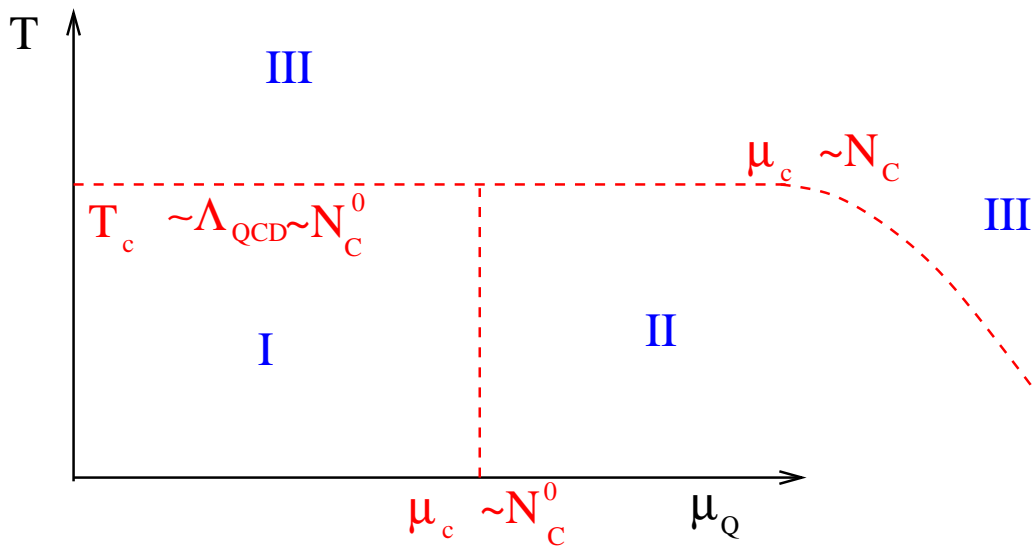}
\caption{(color online)The phase diagram for large $N_c$. See text for a description of the phases I,II,III in various models. \label{phasediag}  See footnote \ref{confusion} about the nomenclature}
\end{figure*}

Perhaps the only relevant quantity with can be uncontroversially be called ``a small parameter'' (albeit not so small in the real world!) is $1/N_c$, where $N_c$ is the number of colors \cite{thooft}.   While the large asymptotically $N_c$ theory shares with QCD asymptotic freedom for hard processes and confinement for soft ones (separated by the energy $\lqcd \sim 250$ MeV, independent of $N_c$) , the  $N_c$ scaling of different observables can be used to establish a model-independent hierarchy.  Thus, the shape of the phase diagram can be said with relative certainty to look like Fig. \ref{phasediag}:   Phases I and III are, respectively, the familiar confined chirally broken Hadron gas (where pressure $\sim N_c^0$) and the deconfined chirally-restored quark-gluon plasma (where pressure $\sim N_c^2$). Since at large $N_c$ gluon loops dominate over quark loops,  the critical \footnote{\label{confusion}Unfortunately the standard nomenclature is somewhat confusing as the subscript ``c'' means ``colors'' in $N_c$ and ''critical'' for thermodynamic quantities (the critical temperature $T_c$,density $\rho_c$, pressure $P_c$, chemical potential $\mu_c$ and so on) } temperature  $T_c \sim N_c^0 \lqcd$, and the critical chemical potential necessary for deconfinement is very high,
$\mu_q \sim N_c^2 \lqcd$.  

Consequently, in the large $N_c$ limit, the phase transition line becomes horizontal for moderate $\mu_q$.  In this limit the transition between zero baryonic density and finite baryonic density matter is infinitely sharp at $N_c \mu_q \sim m_B \sim N_c \lqcd$ \cite{quarkyonic}, since the baryon density $\sim \exp\left[ -N_c\left( \lqcd - \mu_q\right) \right]$ goes to zero exponentially with $N_c$ for chemical potentials less than the baryonic mass.   Thus, a new phase (II) emerges where the nuclear density is $\order{1} \lqcd^3$, parametrically much less  then that required for deconfinement, $\order{N_c} \lqcd^3$.   

Naively, since $\mu_q \sim \lqcd$ is nowhere near the chemical potential required for deconfinement, this phase should just be that of dense nuclear liquid (the large $N_c$ limit of the nuclear liquid, well-studied theoretically and experimentally \cite{liq1,liq2,liq3,liq4,liq5,liq6,liq7}), where nucleons are close to touching each other, yet confinement is still there and degrees of freedom are baryons and mesons.   The critical parameters for this transition, however, are far above what is seen in the real world, in line for the much stronger nuclear force seen at large $N_c$ wrt $N_c=3$ \cite{witten}.
Moreover, as pointed out in \cite{quarkyonic}, at this chemical potential inter-quark distance $\sim 1/N_c$, leading to the apparently paradoxical situation of quarks close enough to interact perturbatively in a confined medium.

\cite{quarkyonic} proposed to solve this conundrum by postulating that in the new phase the quarks below the Fermi surface act as free objects but the Fermi surface excitations are confined.  Thus, while the new phase is confined, the entropy density and pressure feels the quark degrees of freedom and $\sim N_c$, rather than $\sim N_c^0$ as in the usual hadron gas.   This new state of matter, called quarkyonic in \cite{quarkyonic}, should also be realized in our $N_c=3$ world and reachable in heavy ion collisions \cite{quarkyonicfit}  since large $N_c$ is at least qualitatively true in our world\footnote{Recent work has broadened the definition of ``quarkyonic matter'' to other characteristics, in particular related to phenomenology of chiral symmetry breaking and restoration \cite{chiralspiral1,chiralspiral2,sasaki1,sasaki2,sasaki3,hansen,dirkq}.  In this work we use ``quarkyonic matter'' to refer to matter which is confined but whose pressure scales with $N_c$,in accordance to the definition given in \cite{quarkyonic}  }.

A great deal of investigation has gone on to see weather quarkyonic matter appears in any effective theory of QCD.   While a phase transition does seem to exist which has {\em some} of the characteristics described above \cite{sasaki1,sasaki2,sasaki3}, it is not clear weather the most interesting properties ($P \sim N_c$ and chiral symmetry restoration in the confined medium) are physically realized, as we do not have a model realistic enough but still computable.  Other approaches have found no evidence for any such transition \cite{baympaper,philipsenpaper},or have claimed the ``quarkyonic'' phase to have different properties for those claimed in \cite{quarkyonic} (eg \cite{satzquarkplasma} conjectures a chirally broken but deconfined constituent quark plasma).  

As discussed in the introduction, the main difficulty of theoretical investigation in this regime is that there is no reliable approximation technique which is capable of distinguishing between models.
We are thus left with effective theories, such as the NJL and pNJL model \cite{sasaki1,buballa1,buballa2}, or Gribov-Zwanziger confining gluon dynamics \cite{chiralspiral1}.  The results obtained with these models, however, are highly dependent on the assumptions made in them, assumptions which can {\em not} be rigorously shown to derive uniquely from QCD.   In case of the critical point \cite{stephanov}, different models were shown to give very different answers.  Additionally, none of these models contain features unique to non-perturbative QCD, such as exact quark confinement.  As a consequence, the crucial aspect of the quarkyonic hypothesis, scaling of entropy density with $N_c$ in the quarkyonic phase, can not be adequately tested with models such as pNJL \cite{sasaki1,sasaki2}.

A possible way out are techniques deriving from Gauge-string duality \cite{maldacena}.  While no string theory with a dual looking like QCD is known, several models were developed which share with QCD some of its more notable non-perturbative characteristics,such as confinement and chiral symmetry breaking \cite{sugimoto}.    These models can be used to extrapolate to regions inaccessible to pQCD and the lattice, while retaining qualitative aspects of non-perturbative QCD such as its strongly coupled nature and dynamical confinement. 

 A finite chemical potential study \cite{lippert,rozali} within the Sakai-Sugimoto model \cite{sugimoto} has shown that the system looks similar, but is crucially different from \cite{quarkyonic} in several ways.
The basic structure of Fig \ref{phasediag} which emerges is common to \cite{quarkyonic}: The temperature between phases I (confined) and III (deconfined) is insensitive to $N_c$, $T_c \sim N_c^0$.   The curvature of the deconfinement transition line wrt the chemical potential becomes negligible in the large $N_c$ limit (The transition to QGP in the $\mu_B$ plane grows as $N_c^2$). Most intriguingly, a new phase  II emerges, with the transition line at $\mu_q \sim \order{1}  N_c^0 \lqcd$ and the nuclear density as the order parameter, just like in \cite{quarkyonic}.

There are, however, profound differences:  \cite{lippert} finds that both phases I and II are confining and chiral-broken.    No evidence exists that the scaling of the pressure changes between I and II.   In fact, the only difference
between I and II seems to be a discontinuity in the Baryonic density.  The authors of \cite{lippert} interpret phase II as the well studied nuclear gas liquid phase transition \cite{liq0,liq1,liq2,liq3,liq4,liq5,liq6,liq7}, rather than as a new undiscovered phase.   If this interpretation is correct, than searching for the quarkyonic phase and/or the triple point separating I,II,III at upcoming low energy experiments \cite{low1,low2,low3,low4} would be fruitless, as in our $N_c=3$ world the liquid-gas phase has been extensively studied theoretically \cite{liq0,liq1,liq2,liq3,liq4} and pinpointed experimentally \cite{liq5,liq6,liq7}, and its transition line is understood to lie well below $T_c$, so that no triple point exists.

\begin{figure*}
\includegraphics[scale=0.3]{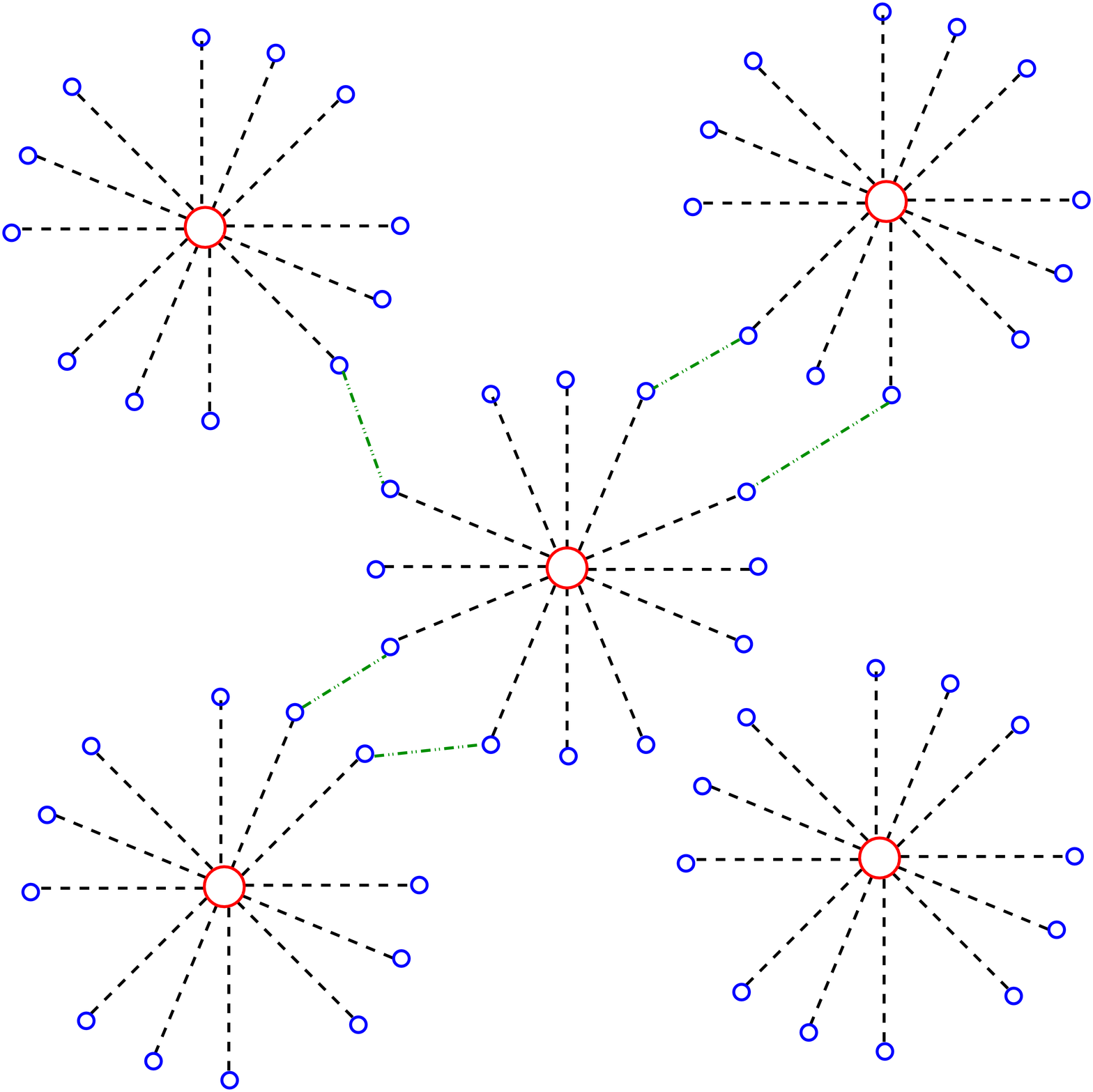}
\caption{(color online)A gas of baryons in the large $N_c$ limit interacting 
\label{ncbinding}}
\end{figure*}

This ambiguity on the phase diagram is compounded by a limitation of our understanding of baryons and their interactions in the large $N_c$ limit.
Baryons at large $N_c$ have long been thought to be well-approximated \cite{witten} by the solitonic ``Hedgehog'' configuration, Fig. \ref{ncbinding}:  A ``flower'' of $N_c$ quarks linked to a Baryon junction via strings (the confining potential).  
The bulk of the baryon mass is carried by the confining strings, which therefore $\sim N_c$, with the proportionality constant given by the quark mass or the string tension.
The baryon radius, on the other hand, $\sim \lqcd^{-1} \sim N_c^0$.

It is generally believed that in the large $N_c$ limit baryons, unlike mesons \cite{thooft} are {\em not}  weakly 
bound states \cite{witten}, but that 2-baryon, 3-baryon and $N$-baryon forces generally $\sim N_c$.   The comparatively weak nuclear force in our world is therefore the result of an accidental cancellation of attractive and repulsive forces.

 Such a scaling,however is somewhat counter-intuitive given the natural hierarchy between multi-body forces arising in effective field theories \cite{bira,weinberg}.   Furthermore, it was found \cite{cohen} that the $N_c$ dependence becomes faster at higher order,\cite{cohen} casting doubt on the existence of a well-defined large $N_c$ limit of nuclear forces.   Recently, the picture has become even more controversial with the conjecture \cite{larry} that the consensus of the skyrmion picture of baryons at large $N_c$ is flawed, and that consequently the nuclear force scales as $N_c^0$.  The rather weak inter-nuclear potential (in comparison to the forces holding quarks and gluons together) observed at $N_c=3$ in this picture would therefore be natural rather than accidental.

This theoretical ambiguity leaves space for exploration with simple effective established models.   Perhaps the oldest and most well-known model capable of capturing the high-temperature dynamics of the non-relativistic liquid-gas phase transition is the Van Der Waals (VdW) equation of state in terms of Pressure $P$, Temperature $T$ and nuclear density $\rho$ . 
This model is undoubtedly extremely rough, and can not be trusted beyond the level of a qualitative
estimate. It does, however, have the virtue of being universal, a mean field expansion general enough to describe
systems as different as water-vapor,nuclear matter \cite{igoreos} (where the potential has been successfully related to effective theories of QCD \cite{virial}) and charged black holes \cite{chamblin1,chamblin2}, the latter generically being
the dual of strongly coupled thermal Gauge theories at finite temperature and chemical potential.

Changing the form of the Baryon-Baryon interactions will not change the form of the Van-Der-Waals terms (or the higher order Virial corrections), but merely change the numerical values of these terms.   Hence, in the vicinity of the phase transition, this model can be trusted to provide a good qualitative description of the relevant physics (such as scaling with $N_c$), though corrections of $\order{1}$ at any given $N_c$, to be calculated by more sophisticated models or on the lattice, would be expected.

We also note that the VdW model has qualitative limitations which prevent it from modeling the ``right-hand side'' of the phase diagram (the dense liquid phase), to check, for example, if in it the pressure would $\sim N_c$ (as proposed in \cite{quarkyonic} for the quarkyonic phase) or $\sim N_c^0$ (as is in our world for the nuclear liquid). This is because in this model interactions are integrated out to a modification of the dispersion relation of nuclei, rather than treated as {\em additional degrees of freedom} that can also contribute to pressure.  This is why the VdW equation of state fails to describe the electron pressure in a crystal, and why it would be inappropriate to describe the quarkyonic phase, where pressure is dominated by partons trapped deep inside the Fermi surface \cite{quarkyonic}.   These limitations, however, do not prevent a qualitative estimate of the phase transition line of the nuclear gas, they just prevent a peek into what happens {\em after} the system crosses that line.   In this work, we shall content ourselves with the first task.

\section{The Van Der Waals nuclear gas at large $N_c$ \label{vdwnc}}
In the large $N_c$ limit, the only $N_c$-invariant scale of the theory is $\lqcd$, the scale at which the 't Hooft coupling constant becomes $\lambda \sim \order{1}$.   While a precise value of this scale depends on the scheme used to calculate it, its roughly $\lqcd \sim N_c^0 \simeq 200-300$ MeV \cite{pdg}.   That this value is very close to the ``constituent'' quark mass, $m_B/N_c \simeq 310$ MeV, is not so surprising given the relationship between confinement and chiral symmetry breaking \cite{casher}:  Assuming that confining effects set in at the scale where $\lambda \geq 1$, and solving for a system of massless fermions in a confining potential (such as a simple harmonic oscillator) will give a ``dynamical mass term'' (or, equivalently, a breaking of chiral symmetry) of the correct order.
Physically, the constituent quark masses will be given by the inverse of the baryon size, which is set by $\lqcd^{-1}$.

Since the only scale in our theory is $\lqcd$,it is natural to expect that any physical quantity is $\sim f(N_c) \lqcd^d$, a dimensionless function of $N_c$ times a power of $\lqcd$ set by the dimensionality $d$ of the quantity.
Henceforward we shall adopt this assumption, and, for brevity, set $\lqcd$ to unity in the equations.    The reader should multiply any dimensionful quantity in the equations by the appropriate power of $\lqcd$ (For example, the Baryon mass is $\sim N_c \lqcd$ in the text, and $\sim N_c$ in the equations).

In this notation, the Van Der Waals parameters $a$, $b$ and the curvature correction become dimensionless $\alpha$,$\beta$,$\gamma$ times the appropriate power of $\lqcd$ (3 for $\alpha$,2 for $\beta$,4 for $\gamma$), and the VdW equation  \cite{chemref2} becomes
\begin{equation}
\left( \rho^{-1} - \alpha \right) \left(  P + \beta \rho^2 - \gamma  \rho^3 \right) = T
\label{vdw}
\end{equation}
Here, $\alpha$ parametrizes the excluded volume and $\beta,\gamma$ the interaction.  When $\alpha,\beta,\gamma=0$ the system reduces to the classical ideal non-relativistic gas.
In the VdW equation usually referred to in textbooks, $\gamma=0$.  This system is solvable more easily, but, as can be noted, at $T=0$ $P$ is generally less than zero.  The next-to-leading order, $\gamma>0$ ``curvature correction'' fixes this artifact, and becomes dominant at low temperatures and high densities.

 The VdW ($\gamma=0$) equation
 can be shown to arise as an effective first order description of {\em any} interacting non-relativistic system where the interacting potential depends on 2-particle interactions only, and the system is dilute, so that the minimum distance between particles is far above the excluded volume.  
Note that a $\gamma$ term arises as a correction to {\em either} of these assumptions, since both 3-body forces and a 2nd term in the Virial expansion can be shown to give rise to $\gamma>0$.  Higher-order corrections behave in a similar way. 

The first assumption of the VdW set-up (2-body interactions dominate) seems to be true in effective field theories\cite{bira,weinberg}, since the strength of an $n-$body interaction typically $\sim (k/\Lambda)^n$ where $k$ is the relevant momentum scale (e.g. the binding energy) and $\Lambda \gg k$ the scale at which the effective theory breaks down. 

 We note that this reasoning is {\em not} in contradiction to \cite{witten}, since its perfectly possible for n-baryon interactions to $\sim N_c (k/\Lambda)^n$.  Thus, 2-body forces are more important than three body forces which are more important than 4-body forces and so on, but {\em all} scale equally with $N_c$.   From this, we learn that generally $\gamma<\beta$ but their $N_c$ dependence should be the same. 


   The second assumption of the VdW set-up  (inter-nuclear distance $\gg$ excluded volume) is certainly true in our world (the inter-nuclear separation is roughly an order of magnitude larger than the nuclear size), but not necessarily in the large $N_c$ world.   To understand why this is the case, we can see Fig. \ref{ncbinding} (in the spirit of the nucleon hedgehog model presented in \cite{witten}): In our world, packing hadrons in such a way that there is a significant overlap area would result in a deconfinement phase transition;  The percolation picture of deconfinement \cite{perc1,perc2} makes this physically intuitive.

In the large $N_c$ world, however, the fraction of quarks which overlap becomes negligible ($\sim N_c^{0}$), while the bulk of the baryon mass remains in the baryon junction.   It is therefore possible for baryons to start overlapping (nuclear density reaches $\sim \lqcd^{3} $) while remaining well-defined objects.

As the number of colors $\rightarrow \infty$, it is therefore logical to postulate $\alpha$ will reach the limit of $\alpha \sim \order{1} \lqcd^{-3}$.   At $N_c$ decreases from $\infty$ to $\order{1}$, $\alpha$ can only increase, so a useful ansatz, universal in any limit where $\alpha$ can be Taylor-expanded, is $\alpha \sim \left(1+ A/N_c \right)\lqcd$.   
To understand the behavior of $A$, it must be remembered that it must sensitive to $N_N$, the number of neighbors a nucleon has in a tightly packed nuclear material.   The more neighbors, the more Pauli blocking of valence quarks must be important, and the more the presence of neighbors will disturb the configuration space part of the quark wavefunction inside the nucleons.
Since,due to confinement, any such disturbance of the nuclear wavefunction adds an energy of $\sim \lqcd$, the nuclear repulsive core will be larger than the inverse of the nuclear separation up to the deconfinement temperature.  If the number of colors is larger than $N_N$, this problem will not exist since it will be possible to arrange quarks so neighbors will be of different colors. In this limit, therefore, baryons can be tightly packed (interbaryonic separation $\sim \lqcd$) without the configuration space part of the baryonic wavefunction being disturbed.

$N_N$, of course, is a function not of $N_c$, but the (fixed) number of dimensions $d$ and ``packing scheme'', $N_N \sim  k(d) N_c^0$. The exact form of the ``kissing number'' function $k(d)$ in arbitrary dimensions is unknown \cite{kissing}, but seems to be approximated by $k(d) \sim 2^{\alpha d}$, with a transcendental $\alpha=0.22...$ .      $k(1,2,3)$ is, respectively, 2,6 and 12.

Therefore, we assume that
\begin{equation}
\label{eqalpha}
\alpha \sim \order{ \frac{N_N}{N_c}} + 1 \sim \order{ \frac{k(d)}{N_c}}+1  \sim \left.  \order{ \frac{10}{N_c}}+1  \right|_{d=3}
\end{equation}
In a sense, the relation above is a parametrization of experimental data combined with natural constraints: We know that the nuclear size can only go as $\sim A +B/N_c$.  We know that in our world the inter-nuclear distance is $\gg \lqcd$, but it can only decrease with $N_c$.   Putting the $\sim N_c^0$ term to $\order{1}$ and the $\sim N_c^{-1}$ term to $\order{10}$ is the only way to account for these limits, as the results of the next section shall show.  The interpretation due to the kissing number is then an appealing physical explanation.

Note that in our world the first term dominates, since for any realistic packing of nuclear matter the number of nearest neighbors is considerably larger than $N_c=3$ (eg for cubic packing, its 6 or 8, depending on weather corners are included. Generally it should go as $k(d)=9$,parametrically larger than 3).   Pauli blocking between valence quarks of neighboring nucleons, therefore, can not be ignored and keeps the inter-nuclear distance parametrically larger than the nucleon size.
In the large $N_c$ but three-dimensional world the second term takes over as the effect of Pauli Blocking becomes negligible, and the excluded volume approaches $\Lambda_{QCD}^{-3}$, since baryons become more tightly packed in the large $N_c$ limit (presumably, the ``jamming'' phase transition described by percolation \cite{perc1,perc2} would be a good description of the liquid-gas transition in this regime).   Note that the small parameter behind the VdW expansion (the scale of the excluded volume over the scale of the inter-nuclear potential) is $<1$ irrespective of $N_c$, even through it reaches $1$ asymptotically with $N_c$.   For a rough qualitative estimate, therefore, the VdW equation is always justified, but we have to keep in mind that higher order quantitative corrections become increasingly important as $N_c \rightarrow \infty$.  These corrections are known to influence the {\em shape} of the phase transition line rather than its basic limits.

The behavior of $\beta$ (and hence $\gamma$) is somewhat ambiguous:  As we remember from \cite{chemref2}, $\beta$ is related to microscopic physics via 
\begin{equation}
\beta = 2 \pi T  \int_{\alpha^{1/3} }^\infty dr r^2 \left( 1- \exp \left[ -\frac{V(r)}{T} \right] \right)
\end{equation}
where $V(r)$ obeys a class of Yukawa-type potentials
\begin{equation}
V(r) \sim \frac{\exp \left[- M r \right]}{r}
\end{equation}
converging to Coloumb as $M \rightarrow 0$.

 As pointed out in \cite{witten}, baryon-baryon interactions by both gluon exchange and meson exchange $\sim N_c$. This would mean that for a low nuclear potential ($\ave{V(x)}/T \ll 1$), $\beta \sim N_c$ It would seem likely, therefore, that the large $N_c$ limit the nuclear ``liquid phase'' is actually a tightly bound ( solid?) material, where the VdW approximation stops holding.
The behavior conjectured in \cite{larry}, on the other hand, would mean that
$\beta \sim N_c^0$ or $\sim \log N_c$, allowing for a nuclear liquid phase not too dissimilar from our world.

In this work, we shall test the consequences of both assumptions, by parametrizing $\beta \sim N_c^\nu$, where $\nu$ can be 0 or 1 ($\beta \sim \log N_c$ has the same qualitative dependence as $N_c^0$, through the convergence rate is parametrically slower).
We shall assume that $\beta$ and $\gamma$ have the same dependence on $N_c$, as expected in both \cite{witten} and \cite{larry}. 
\section{The critical point \label{critpoint}}
From textbook formulae, \cite{chemref2} we can immediately read off the conditions (temperature $T_c$ and density $\rho_c$) of the critical point in case $\gamma=0$.  As argued in the preceding paragraph, the critical point density
\begin{equation}
\rho_c \sim \frac{1}{3 \alpha} \sim  \left( \frac{N_c }{N_N + N_c}  \right)
\label{rhocrit}
\end{equation}
should be $\ll \Lambda_{QCD}^{3}$ in our world, but $\sim \Lambda_{QCD}^3$ in the large $N_c$ world.   

Assuming $\beta \sim N_c^\nu$ yields the following evolution for the critical point temperature
\begin{equation}
T_c \sim \frac{8}{27} \frac{\beta}{\alpha} \sim  \left( \frac{N_c^{1+\nu}}{N_N + N_c} \right) 
\label{tcrit}
\end{equation}
For the case of $\gamma \sim N_c^\nu \ne 0$, the formulae become more complicated, but still analytically tractable, by solving for 
\begin{equation}
\frac{dP}{d\rho} = \frac{d^2 P}{d \rho^2}=0
\end{equation}
with the equation of state being an additional constraint.  We obtain
\begin{equation}
T_c \sim \frac{24 N_c^4+4 N_c^2 N_NF_1  + 2 \sqrt{3} N_c N_N^2 D -3 N_N^3
  F_1 +8 N_c^3 F_2} {288
   (N_c+N_N)^2} N_c^{\nu -2}  \sim N_c^{\nu} g_1 (N_c)
\end{equation}
\begin{equation}
\rho_c \sim \frac{\sqrt{3} \sqrt{8 N_c^2+8 N_c N_N+3
   N_N^2}-3 N_N}{12 (N_c+N_N)} \sim N_c^0 g_2 (N_c)
\end{equation}
where $D=\sqrt{4 N_c^2+4 N_c N_N+3 N_N^2}$,$F_1=\left(\sqrt{3}D-3 N_N\right)$ and $F_2 =\left(\sqrt{3}D+6 N_N\right)$ and $g_{1,2} (N_c)$ are rational functions of $N_c$ where the powers of the numerator and the denominator are the same.

\begin{figure*}
\includegraphics[scale=0.9]{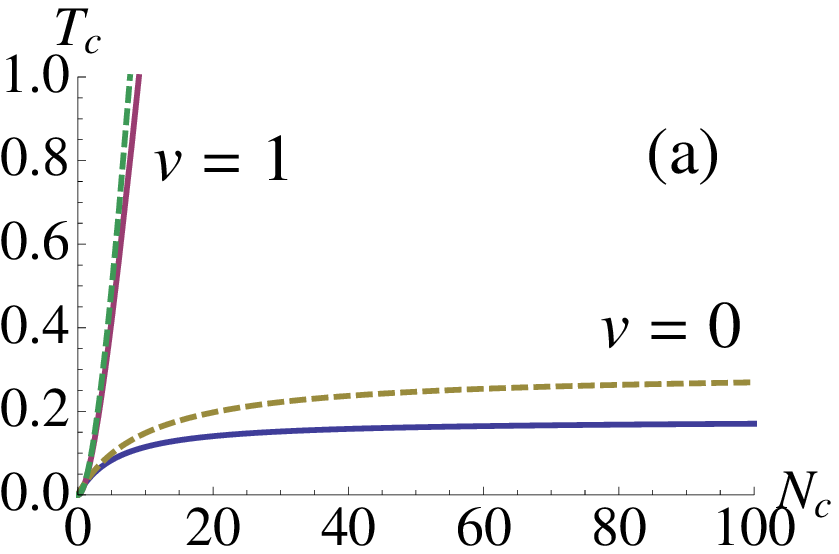}
\hspace{1cm}
\includegraphics[scale=0.9]{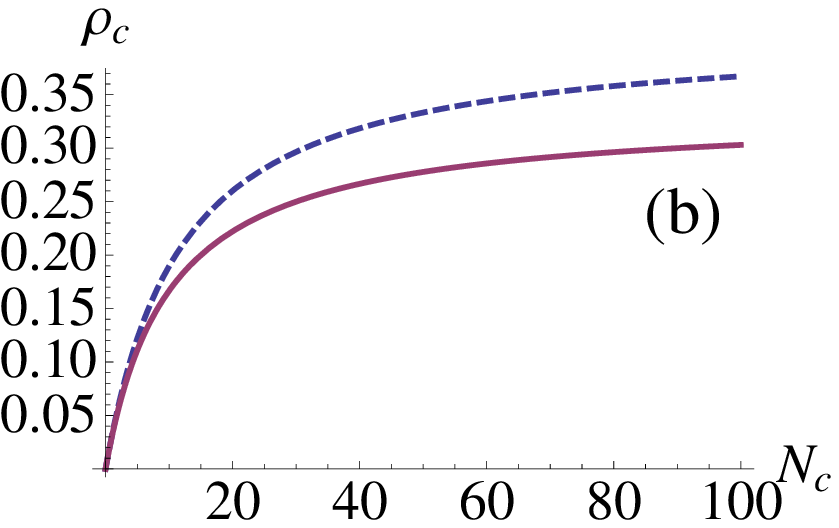}
\caption{(color online)\label{figcrit}The evolution of the critical point temperature (panel (a) ) and density (panel (b) ) with $N_c$ in different scenarios, with $\gamma=0$ (solid) and $\gamma\sim N_c^\nu$ (dashed).  
Note that the density does not depend on $\nu$.  $N_N$ is assumed to be 10, the order of magnitude of the real world value.  Temperature is presented in units of $\Lambda_{QCD}$, while baryon density in units of $\sim 1$ baryon per $fm^{-3}$}
\end{figure*}

The evolution of the $T_c$ and $\rho_c$ are shown in Fig. \ref{figcrit} for $N_N=10$.  As we can see, the presence of $\nu$ makes no qualitative difference for the critical point density $\rho_c$, as generally believed and expected.
$\rho_c$ goes to its asymptotic ``overlapping nuclei'' value regardless of $\nu$.  The same, however, can not be said for $T_c$.

For $\nu=0$ we get the behavior corresponding, physically, to the scenario where the quarkyonic phase and the nuclear matter phase coincide: In the $N_c \ll N_N$ regime, $T_c \ll \lqcd$, while in the large $N_c$ limit it saturates to $\sim \lqcd$ as in \cite{quarkyonic} and \cite{lippert},with a numerical factor of $8/27$ for $\nu=0$ and $\sim 0.31$ for $\nu=1$.
. The numerical factor could very well rise to unity once a more realistic model (including excited nucleons or higher order corrections) is employed, through the deconfinement transition precludes it rising above unity.

For $\nu=1$, $\lim_{N_c \rightarrow \infty} T_c \sim N_c \lqcd$.  Since $\Lambda_{QCD} \sim N_c^0$, this seems to signal that in this picture the
the VdW scenario is inappropriate to describe the nuclear liquid-gas phase transition at large $N_c$, as the critical point would soon overtake the deconfinement temperature already at $N_c \sim 10$, and the binding between nucleons soon overtakes excluded volume effects (As conjectured in \cite{quarkyonic} and revised in \cite{larry}).    
The applicability of VdW would therefore be an accident of our low $N_c$ world, where the binding degrees of freedom do not contribute significantly to the pressure until well above the nuclear liquid-gas phase transition.

Independently of the value of $\nu$, the location of the critical point in the real world, $T \ll \lqcd,\rho \ll \lqcd^3$, can be understood by the realization that we {\em do not} live in a large $N_c$ world, in the sense that $N_c$ is not larger than any other dimensionless scale of the system:  In our world the number of nearest neighbors dominates over $N_c$, lowering the scale at which the ``dense'' (liquid) phase appears to a density much lower to the overlap density of the hadrons.   The evolution of the critical point from $N_c \ll N_N$ to $N_c \gg N_N$ tends to agree with the ``nuclear matter'' interpretation of the new phase as suggested by the calculations in \cite{lippert}, though we can not rule out that this phase also has quarkyonic properties.

In the next few sections we shall demonstrate that this conclusion is valid for the phase diagram shape, and not just for the critical point.

\section{The phase diagram for $\gamma=0$}
\subsection{$T-\rho$ plane}
We now examine the $\beta \sim N_c^0$ case further, by calculating the density jump and the curvature in phase transition space.   For simplicity, we shall concentrate on $\gamma=0$ and leave the low temperature regime for a future work.  As shown in the previous section, away from $T\ll \mu_B$ this approximation should be acceptable.

We rewrite Eq. \ref{vdw} as a cubic equation.
$$\alpha \beta  \rho^3 - \beta  \rho^2 + \rho \left( T + \alpha  P  \right) - P = 0$$
The task of finding the coexistence line of the phase transition diagram is a non-trivial one, giving rise to several computational approaches.
In this work, we have used the parametric solution described in \cite{parametric}.   Given that the entropy difference between the liquid and gas phases is $\Delta s$, the solution can be parametrized by 
\begin{equation}
x_{+,-} = \frac{\alpha }{\rho^{-1}_{g,l}-\alpha } = e^{\pm \Delta s/2} f\left( \frac{\Delta s}{2} \right)
\end{equation}
where, $\rho_{g,l}$ is the density in the gas and liquid phases respectively and
$$f(y) = \frac{y \cosh y - \sinh y}{\sinh y \cosh y -y}$$
The temperature can be found from the same parameters and the requirement that pressure in the liquid and gas phase transitions has to be the same.
solving Eq. \ref{vdw} for pressure, equalizing and doing some algebra gets us
\begin{equation}
T = \left[ \beta  \left( \rho_l - \alpha  \right) \left( \rho_l^{-2} - \rho_g^{-2}  \right)\right]\left[1 - \frac{\rho_l^{-1} - \alpha }{\rho_g^{-1} - \alpha }\right]^{-1}
\end{equation}

\begin{figure*}
\includegraphics[scale=0.7]{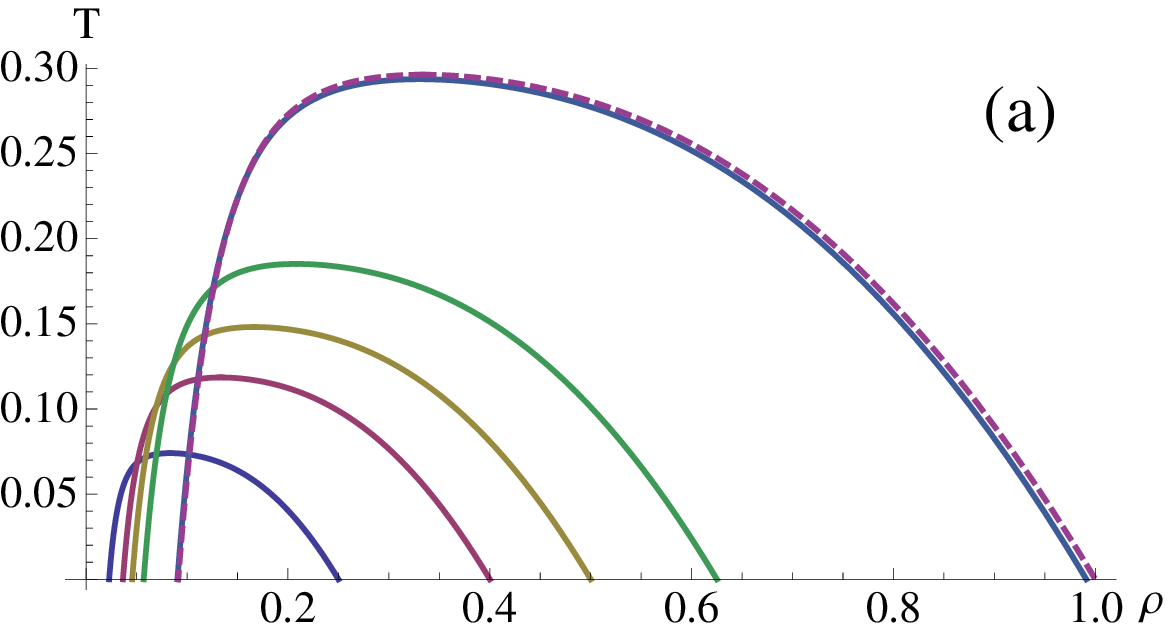}
\includegraphics[scale=0.7]{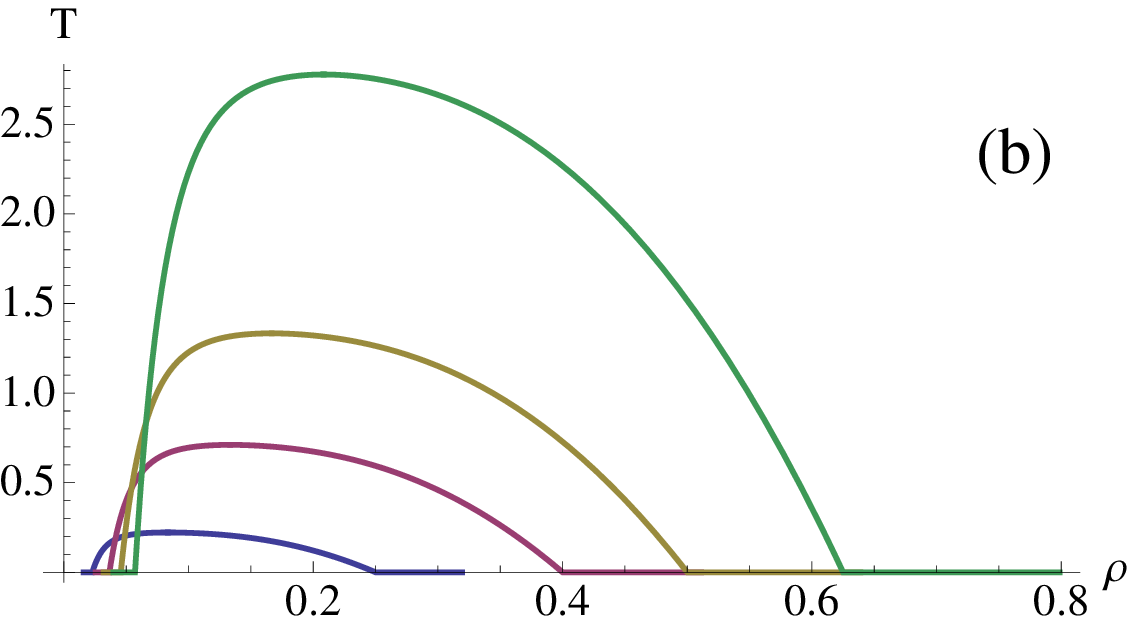}
\caption{(color online)
The phase diagram on the $T-\rho$ plane $\beta \sim N_c^0$ ( panel (a)) and $\beta \sim N_c$ (panel (b)) for 3,5,8,20,100 colors. Higher curves correspond to increasing $N_c$. The dashed line in panel (a) corresponds to the asymptotic limit, absent in panel (b) (As in that case $T_c \sim N_c \Lambda_{QCD}$).  
 $N_N$ is assumed to be 10, the order of magnitude of the real world value. Temperature is presented in units of $\Lambda_{QCD}$, baryon density in units of $\sim 1$ baryon/$fm^3$
\label{nc0} }
\end{figure*}

As can be seen in Fig. \ref{nc0} (panel (a)), as long as the nuclear force potential $V(r) \sim N_c^0$, the VdW transition has a sensible limit.   As $N_c \rightarrow \infty$ ,$T_c \rightarrow \order{1} \Lambda_{QCD}$, $\rho_l - \rho_g \rightarrow 1$ Baryon$\times \lqcd^3 $   
Our world's parameters, where the critical points and the latent heat $\ll \lqcd$ are also qualitatively reproduced, but they lie squarely in the ``low $N_c$'' ($N_c \ll N_N$) region.

As expected, introducing a $\beta \sim N_c$ dependence makes the phase diagram's height shoot up to infinity, while keeping the width constant, an unphysical limit (Fig. \ref{nc0} panel (b))

Finally, we note that at large $N_c$ the mixed phase occupies the bulk of the space between $\rho=0$ and $\rho=\lqcd/3$.  Physically, this is reasonable, since in the diluted phase $\rho \sim \exp\left[-N_c\right] \rightarrow 0$.   Thus, adding ``a few baryons'' into the system  while $\rho \ll \Lambda_{QCD}^3$ will not change the chemical potential.   This indicates the mixed phase should go from $\rho \ll \Lambda_{QCD}^3$ to $\rho \sim \lqcd^3$, exactly as our calculations show.

Our world is, however, very far away from the large $N_c$ limit in this respect, since the nuclear liquid-gas mixed phase actually occupies a small area of the $T-\rho$ plane \cite{liq1,liq3,liq4}.   As Fig. Fig. \ref{nc0} demonstrates, the scaling of the mixed phase is well-accounted for by the interplay between the $N_c \ll N_N$ and $N_c \gg N_N$ limits.

\subsection{$T-\mu$ plane}
The chemical potential can be obtained \cite{dirk,igoreos,chemref2,chemref1} by the textbook thermodynamic relation
$\rho = \left(dP/d\mu  \right)_T$.   Inverting, and writing in terms of $\mu_q=\mu_B/N_c$ we have
\begin{equation}
\label{mubexact}
\mu_q = 1 +\frac{1}{N_c} \left[ \int^\rho_0 f(\rho',T) d\rho' + F(T)  \right]
\end{equation}
where the first term is the nucleon mass and 
\begin{equation}
\label{fdef}
f(\rho,T)= \left( \frac{ dP}{d \rho} \right)_{T} \frac{1}{\rho} = \frac{T}{ \rho  (1-\alpha \rho )^2}+2\beta
\end{equation}
Note that the integral has a logarithmic divergence at $\rho \rightarrow 0$.   This diverge is canceled out in Eq. \ref{mubexact} by $F(T)$, via the requirement that
$\mu_q (T,\rho \rightarrow 0 ) \rightarrow 0$ (So $F(T)=-f(T,\rho \rightarrow 0)$,also a logarithmic divergence).  The resulting $\mu_q$ is always well-defined, and equally valued for $\rho_g$ and $\rho_l$.

The ideal gas limit can be obtained by putting $\alpha=\beta=0$ in Eq. \ref{fdef}.
In this limit, $f(\rho)=T/\rho$ and $F(T)=\ln \lambda^{-3}$ where $\lambda$ is the thermal wavelength $\lambda^{-1}  = \sqrt{m T/2 \pi} \sim \sqrt{N_c T}$.   We therefore recover the ideal gas formula, $\mu_q = 1+ T \ln \left( \lambda^3 \rho \right)$.
Note that,due to the scaling given by Eq. \ref{eqalpha}, $\alpha=0$ is not realized at any $N_c$, since the excluded volume in a confining theory has to $\geq \lqcd^3$. In our world, however, one can generally neglect the excluded volume \cite{dirk} as $\rho^{-1} \gg \alpha$ at the liquid gas phase transition.  This is however {\em not} true at $N_c \gg N_N$.

At low temperatures a correction due to Fermi-Dirac statistics is necessary.  Neglecting the effects of Fermi-Dirac statistics on the excluded volume and interaction, we would get that $\mu_q \rightarrow \mu_q + \Delta \mu_{FD}$, with the correction $\Delta \mu_{FD}$ is given by 
\begin{equation}
\Delta \mu_{FD} = \frac{T}{N_c}\left[ \log(z) - \log \left( \rho \lambda^{3} \right) \right]
\end{equation}
, where the fugacity $z$ can be obtained by solving the the equation linking the density of the ideal gas to the chemical potential \cite{chemref2}
\begin{equation}
\lambda^3 \rho= \frac{4}{\sqrt{\pi}} \int_0^\infty \frac{x^2 d x }{\frac{\exp(x^2)}{z}+ 1 }
\end{equation}
Note that this correction goes to zero in the classical limit.

The inclusion of quantum corrections for only the ideal part of the chemical potential might seem arbitrary, but it is thermodynamically consistent, since, $\Delta \mu_{FD} \rightarrow 0$ as $\rho$ goes to zero.

Note that at large $N_c$ such an exact formula is necessary because the simpler formulae, such as the widely used \cite{igoreos,virial} low temperature formula in terms of the Fermi energy $e_f=(3\pi^2 \rho/2)^{2/3}/(2 m_B) \sim N_c^{-1}$ 
\begin{equation}
\label{mubapprox}
\Delta \mu_{FD} \simeq  \frac{e_f}{N_c}  \left(1 - \frac{\pi^2}{8} \left[ \frac{T}{e_f} \right]^2 \right)
\end{equation}
is {\em not} appropriate.
 If one uses this approximation, the second term of $\mu_q$ will tend to have an $- T^2/(N_c e_f) \sim N_c^0$ contribution which forces $\mu_q$ away from $\Lambda_{QCD}$, in contrast to all other effective phase diagrams \cite{lippert,larry,satzquarkplasma}.  
The problem with this approximation is that it relies on $T \ll e_f$, which, at $N_c \rightarrow \infty$, is not appropriate for any $T$ no matter how low.  This $e_f \rightarrow 0$ scaling is in line with the argument \cite{witten} that
at large $N_c$ baryon is a classical object for which quantum statistics is inappropriate, and furthermore suggests that quantum effects are irrelevant for collective baryon states.  Taking into account the fact that heavy {\em atoms} tend to make a crystalline structure at low temperature (where each atom becomes effectively classical as it is trapped in its location on the crystal), we suggest ( as in \cite{larry,witten,pethick}) an analogous nuclear ground state in the large $N_c$ limit, as it would provide a physical explanation as to why large fermions in the large $N_c$ limit behave like classical objects up to zero temperature.  In our world, of course, the nuclear liquid is very different from a crystal.
 
\begin{figure*}
\includegraphics[scale=0.6]{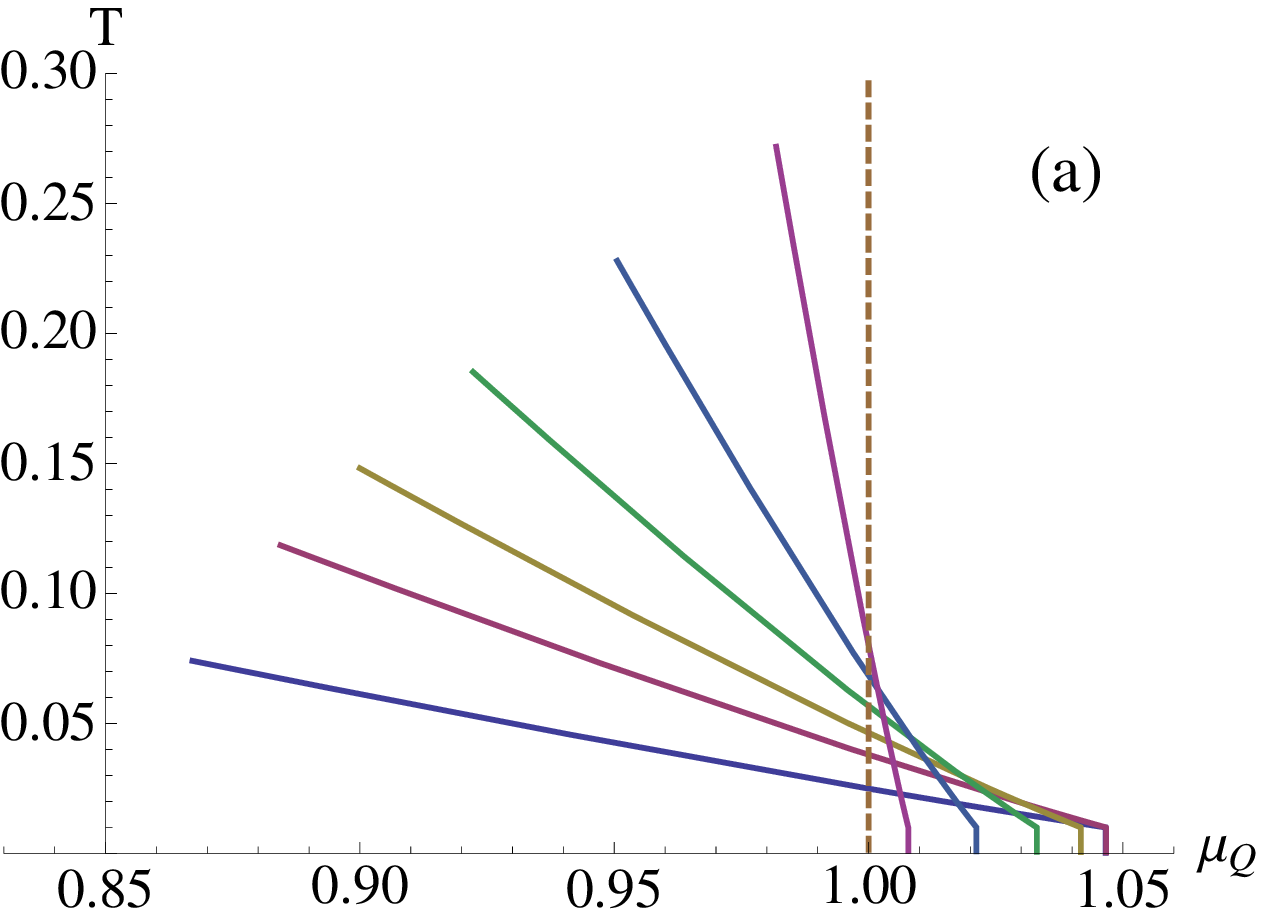}
\hspace{1cm}
\includegraphics[scale=0.6]{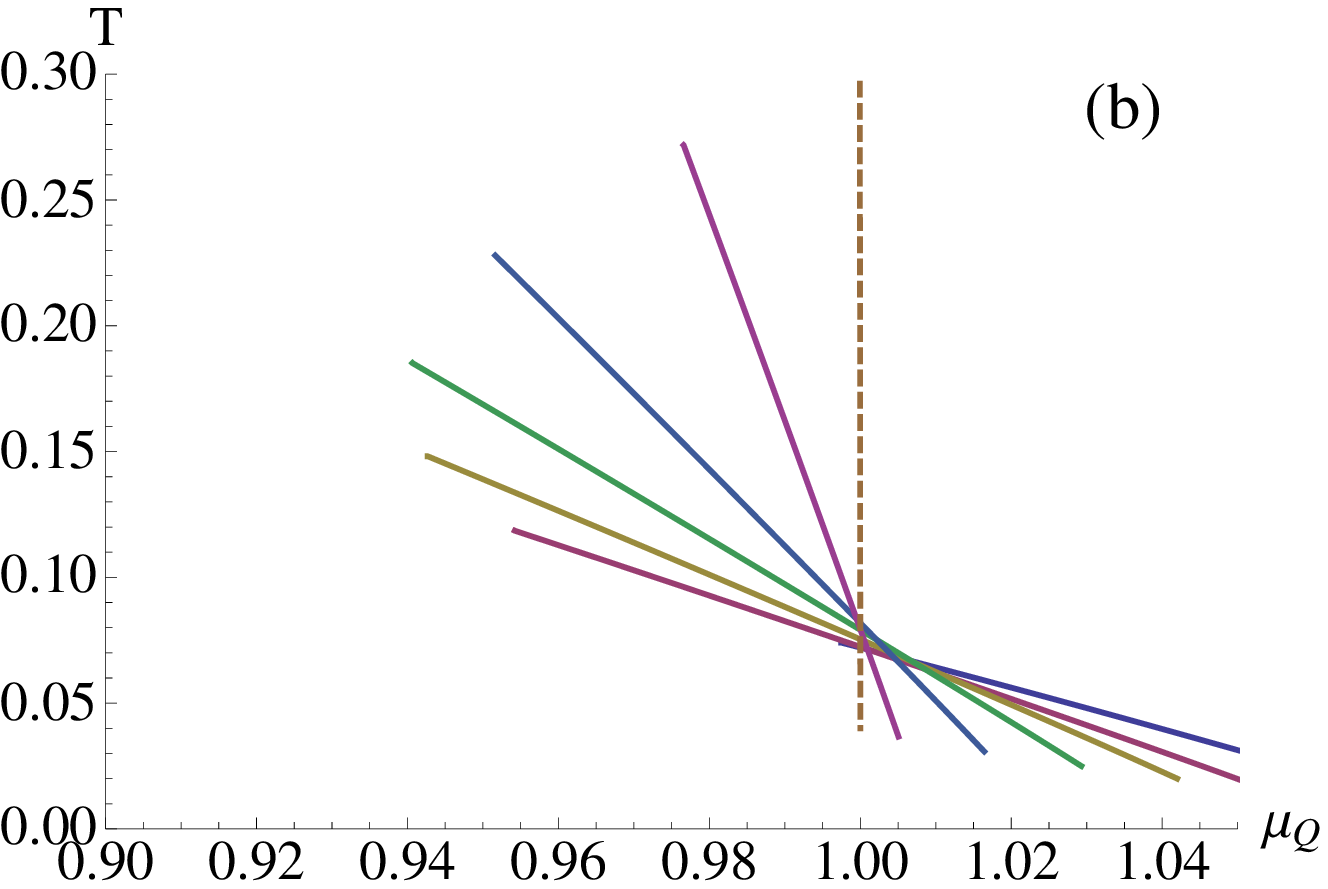}
\caption{(color online)The phase diagram in the $T-\mu_q$ plane for $N_c=3,5,8,10,30,100$, without ( panel (a)) and with (panel (b)) the $\Delta \mu_{FD}$ 
correction.  As $N_c$ increases, the critical points move to the asymptotic limit $N_c=\infty$, represented by the vertical dashed line.  Both $T$
and $\mu_q$ are expressed in units of $\Lambda_{QCD}$. The upper ends of the solid lines correspond to the physical critical points.  The lower ends
of the solid lines correspond to regions where the $\gamma$ and higher terms in the VdW expansion become dominant (Pressure becomes negative unless
$\gamma$ is added), and hence is unphysical.
 \label{ncmu}}
\end{figure*}

The resulting quark chemical potential, including all terms of Eq. \ref{mubexact} and the $N_c$ scaling of $\alpha$ given by Eq. \ref{eqalpha} and $\beta \sim N_c^0$ can be seen in the solid lines of Fig. \ref{ncmu}, without (panel (a)) and with (panel (b)) the $\Delta \mu_{FD}$ correction.
Qualitatively, the diagram looks somewhat different from what we expect the nuclear liquid-gas phase transition to look like, but given the roughness of the model presented here this is not so surprising.   

The fact that this diagram looks very similar to the one obtained in \cite{chamblin1,chamblin2} is encouraging, through one should not
forget that the conserved charges examined in \cite{chamblin1,chamblin2} are very different from the baryonic charge.
The approximately linear shape of the diagram should become more familiar with the inclusion of the $\gamma>0$ curvature term, which would also force the chemical potential to smoothly go to a thermodynamically consistent value at $T=0$ (currently, Eq. \ref{mubexact} stops being well-defined at a lower temperature since, without a $\gamma$-term, $P(T>0)=0$).  

 Note also that,generally $\mu_B>m_B$ at low temperatures, since the excluded volume plays a much larger effect than in our world.   Higher order corrections could again fix this, although it is ultimately due to fact that $\rho$ at the phase transition in our world is $\ll \lqcd^3$, which,as shown in section \ref{vdwnc}, is not natural within large $N_c$ but understandable at low $N_c$.  The inclusion of $\mu_{FD}$ helps bring the phase diagram quantitatively closer to more realistic calculations such as \cite{igoreos}, both at high and low $\mu$, but does not change the qualitative structure of the phase transition line.

\begin{figure*}
\includegraphics[scale=0.8]{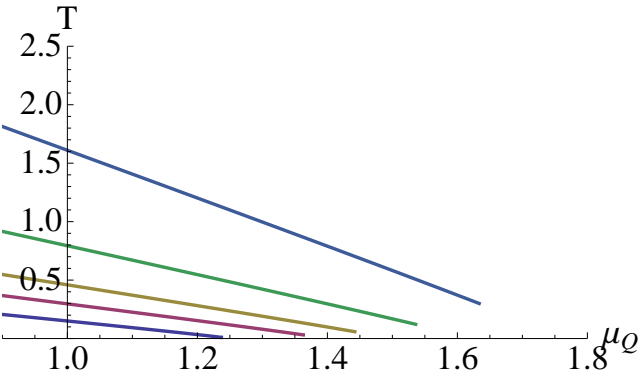}
\caption{(color online) The phase diagram in the $T-\mu_q$ plane for $N_c=3,5,8,10,30$,  assuming $\beta \sim N_c$.
 \label{munc}}
\end{figure*}
The important conclusion to be drown from Fig. \ref{ncmu} is that, while the $N_c \ll \infty$ curves are qualitatively different from realistic nuclear matter and depend on the inclusion of $\gamma$ and higher-order terms, the asymptotic $N_c \rightarrow \infty$ limit is independent of these details, and describes the expected shape of the quarkyonic phase transition: The phase transition line becomes a vertical line in the $\mu_q-T$ plane centered around $\mu_q \sim \Lambda_{QCD}$.   The reason is simply that the baryo-chemical potential $\mu_B$ has a mass component $\sim N_c^1$, and binding energy components $\sim N_c^\nu$.    When $\nu=0$, the latter become negligible, and the chemical potential reaches the limit $\mu_q \sim \Lambda_{QCD} + \order{N_c^{-1}}$ at {\em any} temperature.   When $\nu\geq 1$, as we saw in the previous subsection, the phase diagram does not have a physically plausible large $N_c$ limit.

Fig. \ref{munc} shows what would happen to the $\mu-T$ phase diagram if the nuclear force term $\beta \sim N_c$.
As can be seen, in this case the curvature stays constant at $N_c \rightarrow \infty$.  This is far from surprising, since in this case the binding energy, and hence the curvature term on the phase diagram ($f(\rho',T)$ in Eq \ref{mubexact}) would acquire an $N_c$-linear component.
Fig. \ref{munc},however, also confirms, as the previous sections have shown, that this limit is unphysical at large $N_c$,since both $T$ and $\mu$ go well above $\lqcd$ soon after $N_c=3$
\section{discussion and conclusions}
What are we to make of these findings?   If its really true that $\beta \sim N_c^0$, as suggested in \cite{larry}, we find that the Quarkyonic phase and the liquid-gas phase coincide in the large $N_c$ limit.

The question of whether they also coincide at $N_c \ll \infty$ (ie, if we already ``discovered the quarkyonic phase'' in lower-energy nuclear collisions \cite{liq0,liq1,liq2,liq3,liq4,liq5,liq6,liq7}), or if the liquid-gas transition is distinct from the Quarkyonic phase in this regime, immediately arises.

The fact that only one phase transition, with a structure very similar to the one examined here, arises in the
thermodynamics of charged black holes in AdS spaces \cite{chamblin1,chamblin2}, thought to be dual to the thermodynamics of strongly coupled
Gauge theories of large $N_c$, makes the identification of the quarkyonic transition with the liquid-gas transition
made in \cite{lippert} natural.  One should remember, however, that the charges examined in \cite{chamblin1,chamblin2} (and even in \cite{lippert}) are quite distinct from the
baryonic charge in QCD.

Conceptually, the not-so phenomenologically exciting scenario that the liquid-gas transition {\em is all there is} is plausible from universality arguments: The quarkyonic transition arises from a saturation of entropy by inter-baryonic interactions \cite{quarkyonic}, inter-baryonic interactions are in the Van Der Waals universality class, ergo, the quarkyonic phase is the liquid gas.  

 Furthermore, the question of why does the nuclear liquid look so different from the quarkyonic phase can be sensibly answered by looking at the scaling of $\alpha$:
 The Van Der
Waals terms are “universal” because they arise when a {\em field} giving rise to a nucleon-nucleon interaction 
is integrated out into a nucleon-nucleon interaction {\em term}. Since this {\em field}, in a hot medium, carries entropy, integrating it out under-estimates the entropy of the system. We can now look at Fig. \ref{ncbinding} to get a sense of
how relevant is this undercounting as $N_c$ is varied. If $N_c \ll k(d)$, as in our world, the Pauli exclusion principle keeps the
nuclear excluded volume at a value significantly larger than $\lqcd^{−3}$. In this case, confinement suppresses the exchange
of colored degrees of freedom between the nuclei, so the entropy carried by whatever forces carry nuclear exchange $\sim N_c^0$. It would also mean that the percolating phase transition studied in \cite{perc1,perc2} coincides with deconfinement.

In a $N_c \gg k(d)$ world, however, nuclei touch each other, and colored degrees of freedom can freely percolate
between them. The entropy carried by these percolating degrees of freedom $\sim N_c$, and in the large color limit ends up
overwhelming the total entropy of the system, in much the same way that the electron gas carries most of the entropy
of a metal (Note that the equilibrium entropy of colored objects $\sim N_c$ even if interaction cross-section between these objects is $N_c$-suppressed. The {\em timescale of equilibration} gets longer, but the equilibrium entropy stays the same).   In this limit, the percolation transition \cite{perc1,perc2} does not represent deconfinement but the quarkyonic transition, and the two, in chemical potential, are separated by  $\Delta \mu \sim N_c \lqcd$.
If this scenario is correct, it does indeed imply that looking for the Quarkyonic phase in experiments is pointless.

While the scenario described here is, at the moment, a speculation, this work shows that we {\em do not} live in an $N_c \rightarrow \infty$ world as far as baryonic matter is concerned, since in our world the number of neighbors dominates over the number of valence quarks.  
The transition from a ``low $N_c$ limit'' and the ``high $N_c$ limit'' should happen at $N_c \sim N_N \sim 10$, and theories with $N_c$ below this scale, such as physical QCD, could look qualitatively different from large the $N_c$ calculations.

The physical manifestations of this transition could conceivably be investigated by varying $N_c$ on the lattice at finite chemical potential. Large $N_c$ studies have already been conducted at zero chemical potential \cite{lat1,lat2}, where it was found the thermodynamics is very weakly dependent of $N_c$ already from $N_c=3$.
Our results suggest that, if these studies were extended to finite chemical potential, the results will be very different:   At $N_c \sim 10$ there will be a transition from a thermodynamics much like that of our world, with an order-of-magnitude hierarchy between nuclear matter and tightly packed baryonic (quarkyonic?) matter, to the ``large $N_c$ limit where the two coincide and the liquid-gas critical point temperature and chemical potentials $\sim \lqcd$, and the entropy of the nuclear liquid $\sim N_c$.  

In the strong coupling expansion \cite{fromm}, such a study is already possible with the computational technology available today.  In the weak coupling lattice limit, the sign problem precludes an exact calculation at finite chemical potential.  Currently available algorithms \cite{latmu1,latmu2,latmu3} become very expensive unless $\mu_q/T \ll 1$.  Together with the added numerical cost associated with increasing $N_c$, this means a true test of $N_c$-convergence of lattice QCD at finite baryochemical potential is still a few years away.

This difficulty is, however, computational rather than fundamental:
Since at large $N_c$ the mixed phase goes from very low to very high densities, the lack of convergence with $N_c$ of the critical parameters at finite $\mu$ should be manifest even at relatively small chemical potentials $\mu_q/T \ll 1$ where the approximation techniques described in \cite{latmu1,latmu2,latmu3} can be used.  The computational challenge might be more manageable for 2d lattices, where the ``large'' number of colors is also lower: For the 1D QCD ('t'Hooft model  \cite{hooftmod}) it will be just $N_c=3$ (the fact that any $N_c>2$ is in the infinite limit in this model might account for its relatively trivial phase diagram \cite{hoofttherm}),while for 2D QCD it will be $N_c \sim 6$.

In conclusion, we have discussed the VdW liquid-gas transition in the context of the large $N_c$ scaling of nuclear forces and parameters.
We have determined that the VdW model is unsuitable for describing the large $N_c$ nuclear liquid-gas transition if the nuclear forces scale as $N_c$, as generally accepted after \cite{witten}, but will be suitable if these forces scale as $N_c^0$, as proposed in \cite{larry}.  In the latter case,intriguingly, the large $N_c$ liquid-gas phase transition coincides with the recently proposed Quarkyonic transition \cite{quarkyonic}, although the two are well-distinct at $N_c=3$.  This hierarchy is due to the interplay between two relevant scales, the number of colors $N_c$ and the number of nuclear neighbors $N_N$, with the second scale larger than the first in the real world but smaller in the large $N_c$ world.   A lot of further work, both theoretical (lattice simulations, effective theory models, etc) and experimental (low energy scans) is required to interpret these results in a context closer to fundamental QCD.

IM acknowledges support provided by the DFG grant 436RUS 113/711/0-2 (Germany)
and grant NS-7235.2010.2 (Russia).
G.T. acknowledges the financial support received from the Helmholtz International
Center for FAIR within the framework of the LOEWE program
(Landesoffensive zur Entwicklung Wissenschaftlich-\"Okonomischer
Exzellenz) launched by the State of Hesse.  G.~T. thanks M.~Gyulassy and Columbia
University for the hospitality provided when part of this work was done. The authors thank Jorge Noronha, Larry McLerran,Francesco Giacosa, Matthew Lippert, Chihiro Sasaki, Daniel Fernandez-Fraile and Hovhannes Grigoryan for discussions.

\end{document}